\documentclass[runningheads]{llncs}
\usepackage[T1]{fontenc}
\usepackage{graphicx,verbatim}
\usepackage{todonotes} % Remove for submission
\usepackage{amsmath,amssymb}
\usepackage{bm}
\usepackage{cleveref}
\usepackage[font=small,labelfont=bf,labelsep=period]{subcaption}
\usepackage[dvipsnames]{xcolor}

\newcommand{\notebox}[1]{ }
\newcommand{\noteboxdone}[1]{ }

\begin{document}
%
% \title{Lesion Skin Image Augmentation for Fairness
%   through Skin Colour Disentangling Model}
\title{Automated Disentangling Analysis of Skin Colour for Lesion Images}
\titlerunning{Automated Disentangling Colour Analysis of Skin Images}
\author{Wenbo Yang\orcidID{0009-0000-7528-1925} \and
Eman Rezk\orcidID{0000-0002-2531-0799} \and
Walaa M. Moursi\orcidID{0000-0002-0113-9309} \and
Zhou Wang\orcidID{0000-0003-4413-4441}
%\inst{1}
\authorrunning{F. Author et al.}
\institute{University of Waterloo, Waterloo, ON, Canada, N2L 3G1} \\
%\email{{w243yang,e2rezk,walaa.moursi,z70wang}@uwaterloo.ca} % \\
\email{\{w243yang,e2rezk,walaa.moursi,z70wang\}@uwaterloo.ca} % \\
}
  
\maketitle              % typeset the header of the contribution
\begin{abstract}
%Machine-learning models working on skin images often have degraded performance when the skin colour captured in images (SCCI) differs between training and deployment. Such differences arise from entangled environmental factors (e.g., illumination, camera settings), and intrinsic factors (e.g., skin tone) that cannot be accurately described by a single ``skin tone'' scalar.
Machine-learning models applied to skin images often have degraded performance when the skin colour captured in images (SCCI) differs between training and deployment. These discrepancies arise from a combination of entangled environmental factors (e.g., illumination, camera settings) and intrinsic factors (e.g., skin tone) that cannot be accurately described by a single ``skin tone'' scalar -- a simplification commonly adopted by prior work.
To mitigate such colour mismatches, we propose a skin-colour disentangling framework that adapts disentanglement-by-compression to learn a structured, manipulable latent space for SCCI from unlabelled dermatology images.
To prevent information leakage that hinders proper learning of dark colour features, we introduce a randomized, mostly monotonic decolourization mapping. To suppress unintended colour shifts of localized patterns (e.g., ink marks, scars) during colour manipulation, we further propose a geometry-aligned post-processing step.
Together, these components enable faithful counterfactual editing and
answering an essential question: ``What would this skin condition look
like under a different SCCI?'', as well as 
% The learned latent space supports 
direct colour transfer between images and controlled traversal along physically meaningful directions (e.g., blood perfusion, camera white balance), enabling 
educational visualization of skin conditions under varying SCCI. We demonstrate that dataset-level augmentation and colour normalization based on our framework achieve competitive lesion classification performance.
Ultimately, our work promotes equitable diagnosis through creating diverse training datasets that include 
different skin tones and image-capturing conditions. 
\keywords{Skin Colour Disentanglement \and Counterfactual Image Synthesis \and Data Augmentation}
% Authors must provide keywords and are not allowed to remove this Keyword section.

\end{abstract}

\section{Introduction}

\begin{figure}[t]
  \centering
  \includegraphics[width=0.95\textwidth]{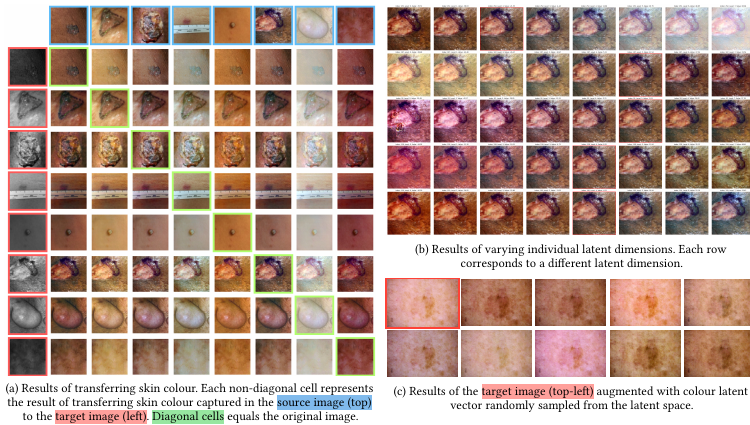}
  \caption{
    Our model is capable of changing the captured skin colour after
    training.
    % and the
    % correlation matrix of the latent entries is shown in (e)
  }
  \label{fig:great}
\end{figure}
\begin{figure}[t]
  \centering
  \includegraphics[width=0.95\textwidth]{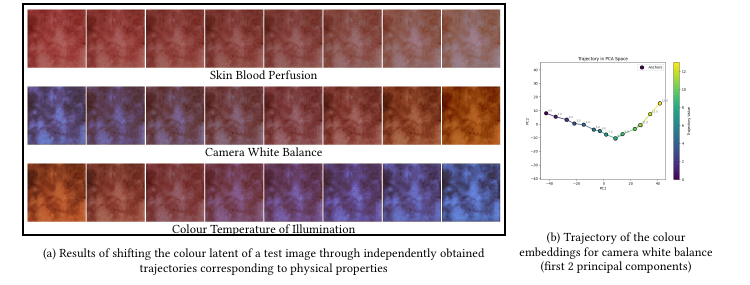}
  \caption{
    % Results of shifting an image along the trajectory of
    % different physical factors in the latent space (a, b, c). 
    % The PCA plot of the trajectory (b) is shown in (d).
    Trajectories with physical meanings can be found in the latent
    space.
  }
  \label{fig:latent-physical}
\end{figure}

Computer-aided dermatology diagnosis systems have improved
significantly in recent years. However, similar to human experts, their performance
degrades when training and deployment data exhibit
mismatched skin colours captured in images (SCCI).
Benmalek et al.\ \cite{benmalek2024skin} reports accuracy gaps reaching
10\%--30\%
between light and dark skin presentations.
The mismatch of SCCI is caused by multiple entangled factors, including
intrinsic subject properties (e.g., skin tone), and
environmental factors (e.g., illumination, camera white balance, and
other device settings) \cite{howard2021reliability,weir2025evaluating}.

% Two categories of methods have been proposed to mitigate the issue of
% colour mismatch.  
Existing mitigation strategies for colour mismatch fall into two
categories. 
First, most recent works in dermatology focus on augmenting
the training set using augmented images  \cite{aggarwal2022ai} or 
partially or fully synthetic images \cite{eman22,dermgan}.  While these methods show
promising performance improvement, they only focus on intrinsic
skin-tone change and 
usually consider skin
colour on individual images without a structured colour model, offering limited
control of the result and distribution matching of datasets.  The
single-value scalar or categorical skin-tone descriptors
(such as FST \cite{fitzpatrick} and MST \cite{monk}) used by these methods
have been shown to have very poor correlation with the skin colour shown
in photos \cite{howard2021reliability,weir2025evaluating}, because 
it does not reflect other factors (including blood perfusion
and environmental factors). 
Second, colour normalization
aims at mitigating the
mismatch by converting each image to a ``standard'' colour. 
While such a method has been widely used in histology
\cite{cong2021,xu2025stain,tosta2023}, 
only limited works have been proposed for skin images
\cite{iyatomi2011,schaefer2011,dermoccgan}, possibly due to the more complex
colour formation process.
Currently, these methods focus on environmental factors 
without explicitly considering the intrinsic factors that
affect the SCCI. 
% Since the factors affecting SCCI are entangled and difficult to measure,
% we propose a tool 
%
Moreover, educators and future physicians need
a tool for visualizing the same skin condition under
different SCCI, which requires a representation that can be
traversed in a controlled and interpretable way.

In this work, we propose an unsupervisedly learned skin-colour
disentangling framework that treats SCCI as the outcome of
multiple entangled factors.
% Building on the information bottleneck (IB) principle
% \cite{ib}, which has been recently extended to low-level vision
% \cite{li2025bitrate,yang2025},
Enabled by recent applications of the information bottleneck principle
\cite{ib} to low-level vision feature disentanglement \cite{li2025bitrate,yang2025},
our method adapts \cite{yang2025}'s disentanglement-by-compression
framework and compresses skin colour into an organized latent
space whose entries are approximately independent and can be
individually adjusted.
Faithful disentanglement of SCCI possesses unique challenges compared to
\cite{yang2025}. 
First, to avoid shortcuts that leak skin-darkness/shading information and
obstruct the representation learning, we introduce a randomized,
mostly monotonic decolourization mapping that generates the
colourless input required by the framework.
Second, to suppress the unintended colour shift of localized patterns (e.g., ink marks,
scars), we propose a geometry-aligned post-processing step to
selectively reject unreliable changes.
The ability of faithful controlled manipulation in a latent space enables
visualizing the lesion under a
different SCCI.
%\todo{this sentence is repeated in the contribution list, maybe we can
%remove it from there}

After training, SCCI can be transferred directly from source
images to targets (\cref{fig:great}(a)), and
individual latent entries can be adjusted independently
(\cref{fig:great}(b)).
The proposed method improves health equity by adjusting the training set for different populations. 
Dataset-level augmentation can be performed by sampling
from a target distribution (\cref{fig:great}(c)), and
data normalization can be performed by setting the colour embedding to a
standard value.
Physically meaningful trajectories can be identified and traversed in the latent
space (\cref{fig:latent-physical})  for educational visualization.
In summary, the main contributions of this work are:
\begin{itemize}
  \item An unsupervised model that answers the counterfactual
    \emph{``What would the skin condition look like if the
    skin colour captured in the image were different?''} via latent-space manipulation
    and transfer, enabled by the randomized decolourization module we
    propose.
  \item A geometry-aligned post-processing step that corrects
    unintended colour shifts,  improving faithfulness of the intended appearance transformation.
  \item Augmentation and normalization strategies for
    downstream tasks using the proposed SCCI modelling method.
\end{itemize}

\section{Methodology}

\subsection{Model Architecture for Perceived Skin Colour Modelling}

\begin{figure}[t]
  \centering
  \includegraphics[width=0.95\columnwidth]{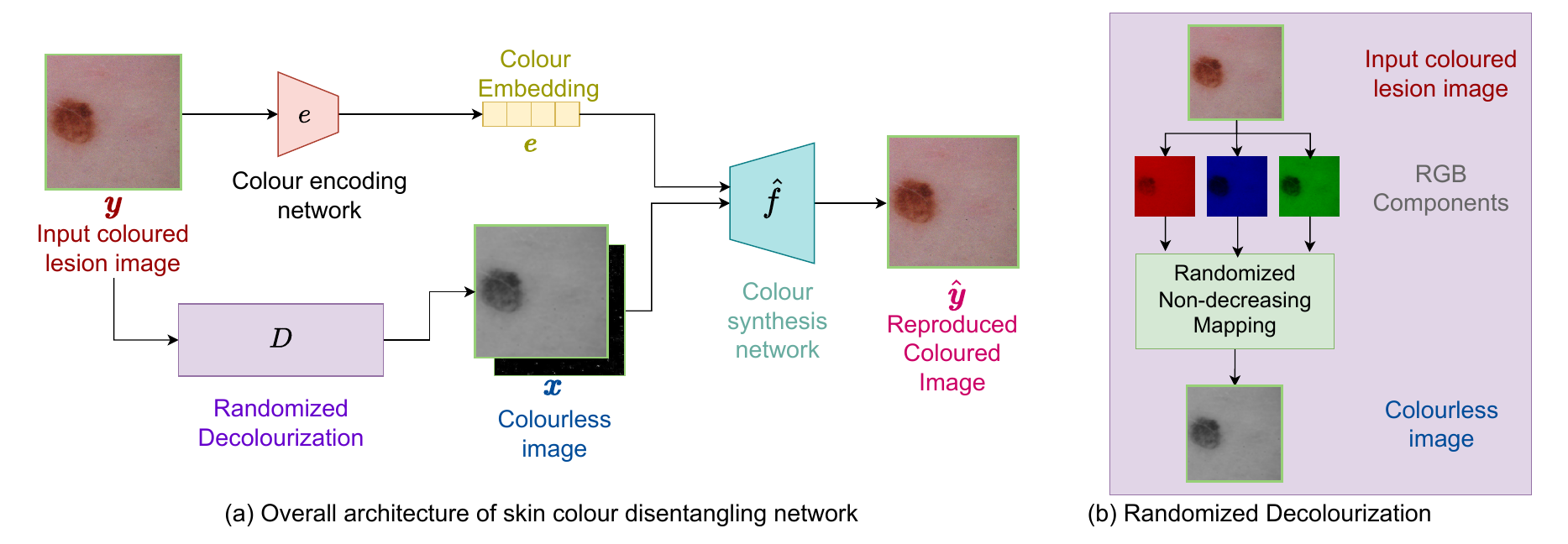}
  \caption{Architecture of the skin colour disentangling model.}
  \label{img:arch}
\end{figure}

Since perceived skin colour captured in images (SCCI) is 
influenced by multiple entangled factors (e.g., skin tone, illumination, and camera
settings), we model it with a disentanglement-by-compression framework adapted
from \cite{yang2025} (retaining most of its architecture).  As shown in \cref{img:arch}(a), the colour encoding network
\(e\) maps an input image \(\bm y\) to a low-bitrate colour embedding
\(\bm e = e(\bm y)\).  A colour synthesis network \(f\) then reconstructs
\(\hat{\bm y} = f(\bm x, \bm e)\) from \(\bm e\) and a colourless image
\(\bm x\) that preserves geometry but suppresses colour from \(\bm y\).  By
constraining \(\bm e\) to be just sufficient for faithful reconstruction, \(e\)
is encouraged to retain primarily global SCCI information.  
At test time,
replacing \(\bm x\) with another image enables colour transfer, while
editing or sampling entries of \(\bm e\) enables controlled colour
manipulation.

% A key difference from \cite{yang2025} is that the colourless image \(\bm
% x\) is generated from \(\bm y\) through a decolourization module
% \(g_{\bm \alpha}\), which will be detailed in the remainder of this
% section.
% We shall detail the decolourization module \(g_{\bm \alpha}\), 
% which removes the colour information from \(\bm y\) to generate \(\bm
% x\),
% \todo{Update the figure, replace \(D\) with \(g_{\bm \alpha}\)}
% in the remainder of this section. 

% \subsection{Image Decolourization}

% Original: 13.3 lines (excluding the equations)

% 10 lines (excluding the equations)
\subsubsection{Decolourization}
Unlike \cite{yang2025}, which assumes a separately provided \(\bm x\) for
each \(\bm y\), we propose
a randomized decolourization module to 
generate \(\bm x\) from \(\bm y\). 
A fixed linear greyscale conversion can leak
luminance cues correlated with skin darkness and shading, which weakens
control over perceived darkness. 
We therefore use a
randomized pixel-wise mapping \(g_{\bm \alpha} : [0,1]^3 \to [0,1]\) with
fixed endpoints
\begin{equation}
  g_{\bm \alpha}(\bm 0)=0, \qquad g_{\bm \alpha}(\bm 1)=1,
\end{equation}
where \(\bm \alpha\) is resampled for each training image in
each epoch, and \(g_{\bm \alpha}\) is chosen to be mostly increasing in
each channel.  In practice,
a linear combination of monotonic quadratic terms is used to balance simplicity and
diversity in concavity:
\begin{equation}
  g_{\bm \alpha}(\bm c)
  :=
  \sum_{i=1}^3 \sum_{j=i}^3 \alpha_{ij} c_i c_j
  +
  \sum_{i=1}^3 \sum_{j=i}^3 \beta_{ij} (1 - (1 - c_i)(1 - c_j)),
\end{equation}
where \(\bm c=(c_1,c_2,c_3)\) is the RGB vector and
\(\bm \alpha=(\alpha_{ij},\beta_{ij})_{i\le j}\) is sampled such that
\(\sum_{i\le j} (\alpha_{ij}+\beta_{ij})=1\) and
\(\alpha_{ij},\beta_{ij}\ge -\epsilon\) for a small \(\epsilon\) that is
chosen to balance flexibility and monotonicity.

% \subsection{Image Decolourization}

\subsection{Post-Processing for Colour Correction}
Because the model is designed to capture SCCI (a global
feature across the entire image),
localized colour patterns
(e.g., ink marks, scars, irritations) may undergo undesirable colour
shifts during transfer or sampling. 
We correct the colour of such regions by defining a post-processing operator \(P\)
that suppresses changes at pixels that were already hard to reconstruct.
Let \(\hat{\bm y}=f(\bm x,\bm e)\) with
\(\bm e=e(\bm y)\), and let \(\hat{\bm y}'=f(\bm x,\bm e')\) be the
manipulated output, where \(\bm e'\) is obtained
from another image, sampling, or manual adjustment.
We define a rejection weight \(\bm w \) by
\begin{equation}
  w_{ij} := h(d_{ij}), \qquad
  \text{where }
  d_{ij} := \frac{1}{1-\tau}
  \max\!\left(\|y_{ij}-\hat{y}_{ij}\|-\tau,\,0\right),
\end{equation}
\(\tau\) is a small tolerance threshold,
and \(h:[0,1]\to[0,1]\) is monotone increasing.
The post-processing operator \(P\) only accepts the 
change proposed by \(f\) when the reconstruction error \(d_{ij}\) is
small: 
% \(\bm y^o=P(\hat{\bm y}')\) is defined as
% \[
% y^o_{ij}:=y_{ij}+(1-w_{ij})(\hat{y}'_{ij}-y_{ij}).
% \]
\begin{equation}
  P(\hat{\bm y}') := \bm y + (1-\bm w) \odot (\hat{\bm y}' - \bm y).
\end{equation}

\subsection{Data Augmentation and Normalization for Downstream Tasks}

A typical downstream ML model is trained on skin images with distribution
\(S\) and tested on \(T\).  Labels for \(S\) are always available,
whereas labels for \(T\) are not during training.
In some scenarios, however, unlabelled examples from \(T\) are
available in advance
(e.g.\ when a clinical site has collected some patient images but
diagnosis is not yet available).%\todo{better expression?}
It is well known that downstream performance can degrade when the
SCCI distributions of \(S\) and \(T\) mismatch.

\subsubsection{Training Data Augmentation} 
In physician education, exposure to skin colours in the population that they
will serve is essential for building an accurate mental model. 
Similarly, a ML model can benefit from a training set augmented
to match the SCCI distribution of a target cohort:
\begin{equation}
  S_{\text{aug}} := \left\{
    P \circ f \left(
    g_{\bm \alpha}(\bm y^{(i)}), \bm e^{(i,k)}
    \right)
    :
    \bm y^{(i)} \sim S,
    \bm e^{(i,k)} \sim e(T)
  \right\}
  \label{equ:aug}
\end{equation}
where 
\(f\) is trained on \(S \cup T\) without labels and
\(\bm e^{(i,k)} \sim e(T)\) is sampled from the embedding distribution of
the test set. 
Since it is needed to model the SCCI distribution of \(T\), some
unlabelled images from \(T\) are needed during training.
This mode suits deployment settings where unlabelled target-domain
images are available before testing
while diagnostic labels remain unavailable.
This procedure only modifies the training set, 
with no change needed to the test-time procedure,
and can be used to generate training examples for physician education.

% \subsubsection{Colour Normalization}
% In contrast, colour normalization 
% is more ML-centric.  It
% aims to adjust each training and test
% image to a ``standard'' colour:
% \begin{equation}
%   S_{\text{nrm}} := \left\{ 
%     f_p\left( 
%     g_{\bm \alpha}(\bm y^{(i)}), \bar{\bm e}
%     \right)
%   \right\}_{i \in \mathbb{N}_N}
%     {\text{ and }} ~
%   T_{\text{nrm}} := \left\{ 
%     f_p\left( 
%     g_{\bm \alpha}(\bm y^{(i)}), \bar{\bm e}
%     \right)
%   \right\}_{i \in \mathbb{N}_N}, 
%   \label{equ:norm}
% \end{equation}
% where \(\bar{\bm e}\) is mean colour embedding of the training set. 
% Compared to training data augmentation, colour normalization can be
% used for online testing, 
% where each test image can be individually
% tested without knowing a large set of test images in advance.
% This is possible because
% \(f\) only needs to be trained on \(S\), 
% \todo{however, we are also going to show the data for training on \(S
% \cup T\)}
% and there is no need to know the distribution of \(T\) in advance. 
% Still, this method needs a modified and more computationally expensive test-time
% procedure than training data augmentation.

\subsubsection{Colour Normalization}
In contrast, colour normalization is more ML-centric.
It maps every image to a ``standard'' colour, reducing variability
across devices and acquisition conditions in both training and
evaluation:
\begin{equation}
  \begin{aligned}
    S_{\text{nrm}} & := \left\{ 
    P \circ f\left( 
    g_{\bm \alpha}(\bm y^{(i)}), \bar{\bm e}
    \right)
    :
    \bm y^{(i)} \sim S
  \right\}
  , \text{ and } \\
      T_{\text{nrm}} & := \left\{ 
    P \circ f\left( 
    g_{\bm \alpha}(\bm y^{(i)}), \bar{\bm e}
    \right)
    :
    \bm y^{(i)} \sim T
  \right\}
  \end{aligned}
  \label{equ:norm}
\end{equation}
where \(\bar{\bm e}\) is the mean colour embedding 
of the training
set
used as the
``standard''.
Unlike augmentation, normalization supports online use
(at the cost of running \(f\) at test time of downstream tasks): each test
image is processed independently without knowledge of \(T\)'s
distribution (and without needing examples from \(T\) during training), and \(f\) can be trained on \(S\) alone.

\section{Experiments and Results}

%\subsection{Skin Colour Model Training}
%\todo{Maybe we need to retrian the baseline model without using edge
%images}

% Each model is trained on a H100 GPU for 50 epochs or 24 hours (whichever
% is earlier) using the same loss functions, training procedure, and
% optimization settings as in \cite{yang2025} (except we adjusted 
% \(\lambda_{\text{bpp\_g}} = 0.003, \lambda_{\text{diver}} = 0.05\),
% and 
% \(\lambda_{\text{color}} = 0.1\)). 
% For different downstream tasks, different training set might be used.
% For demonstrative and qualitative experiments, the model trained for
% dataset augmentation is used, which is trained on the training and test
% sets of Rezl et.\ al.\cite{eman22} and ISIC-2020 \cite{isic2020} (both without labels). 
% In the post-processed process, we use \(\tau=0.1\) and \(h(d) = d^2\).
% The code of the project will be released upon publication.

Each model is trained on an H100 GPU for 50 epochs 
% or 24 hours (whichever comes first) 
using the same loss functions, training
procedure, and optimization settings as in~\cite{yang2025},
except that we set
\(\lambda_{\text{bpp\_g}} = 0.003\),
\(\lambda_{\text{diver}} = 0.05\), and
\(\lambda_{\text{color}} = 0.1\).
Different downstream tasks may use different training sets.
For qualitative experiments, we use the augmentation model
trained on the training and test sets in ~\cite{eman22} and ISIC-2020~\cite{isic2020}
(both without labels).
For post-processing, we set \(\tau=0.1\) and \(h(d) = d^2\).
The code will be released upon acceptance.

\subsection{Qualitative Results of Skin Colour Matching and Sampling}

\notebox{
The videos showing the results of VARYING skin colour really warrant the
multimedia supplementary material. 

Additional legitimate materials to put in the supplementary materials:
\begin{itemize}
  \item Some images from the augmented training set
  \item Larger figures showing the qualitative results
\end{itemize}
}

% \begin{figure}[t]
%   \centering
%   \begin{subfigure}{0.48\textwidth}
%     \includegraphics[width=0.9\textwidth]{imgs/demos/varied_latents_image_5.png}
%     \caption{By varying individual latent entries}
%     %\label{fig:first}
%   \end{subfigure}
%   \hfill
%   \begin{subfigure}{0.48\textwidth}
%     \includegraphics[width=\textwidth]{imgs/demos/examples/plot.pdf}
%     \caption{By random sampling of the latent vector}
%     %\label{fig:second}
%   \end{subfigure}
%   \caption{Results of varying skin colour features through latent
%     space manipulation. Each row of (a) shows the results of varying an
%     entry in the latent vector, while (b) shows the results of randomly
%     sampling the entire latent vector.
%   }
%   \label{fig:qualitative-manipulation}
% \end{figure}

We selected several challenging examples from
SkinCap~\cite{skincap} and ISIC-2020~\cite{isic2020} as test
cases and generated results by transferring the SCCI
between them.
As can be seen in \cref{fig:great}(a), our model effectively transfers
the SCCI without altering the skin structure or lesion
appearance.

By computing the variance of each entry in the
256-dimensional latent vector \(\bm e\), we identified a
small number of active entries.
The results of 
individually varying the first five active entries are shown in
\cref{fig:great}(b), each shows a semi-interpretable physical
factor.
In addition, we independently sample each entry of \(\bm e\)
from the empirical marginal distribution of the SkinCap test set to augment a
training image.
The results are shown in \cref{fig:great}(c).
The figure confirms that the model produces diverse yet
realistic skin colour variations.

\subsection{Exploration of the Latent Space}

\notebox{
\begin{itemize}
  \item Show the two curves matching REAL physical meanings (blood
    circulation and white balance of camera)
  \item (If have space) 
  \item (If \textit{really} have space) show error maps
\end{itemize}
}
% generated by the script run_scripts/pc32/run_pc32_test-7-local-new-stat-plots.sh

Realistic physical meaning can be observed in the latent space.
We captured a sequence of trajectory-building sample photos
(TBSPs) corresponding to three physical factors:
(a) skin blood perfusion (via temperature change),
(b) camera white balance settings, and
(c) the colour temperature of lights.
For each factor, the TBSPs are fed into the colour encoding
network to produce a sequence of latent vectors, which are
connected in a piecewise-linear manner to form a sample
trajectory (STra) in the latent space.
To visualise the effect, we randomly select images and
vary their colour latent \(\bm e\) along a curve parallel to
the STra.
Results from one test image are shown in
\cref{fig:latent-physical}(a,b,c)
(animated examples can be found in the
supplementary materials).
The PCA plot of the STra for camera white balance settings
(\cref{fig:latent-physical}(d)) traces a smooth curve,
indicating the model has learned a continuous representation
of this physical factor.

\subsection{Dataset Augmentation and Colour Normalization for Lesion
Classification}

\notebox{
Emphasize that the main purpose is not to propose a SOTA classification model, but to explore the potential of the augmentation method and compare its effectiveness to existing ones. 

In addition, also discuss the colour normalization results

Emphasize colour normalization and training set augmentation have
different applicable fields for different constraints, although they can
be demonstrated through similar experiments.
\begin{itemize}
  \item augmentation does not require putting test images to the model,
    easier deployment
  \item normalization: colour model does not need to see the test set
    (which might be desirable for regulations pertaining privacy)
\end{itemize}
}

\begin{table}[t]
  \fontsize{8pt}{9pt}\selectfont
  \centering
  \caption{Accuracy of lesion malignancy classification
    models trained on different augmented or normalized training sets.
    Results of existing methods are obtained from the original authors.
  }
    \label{tbl:classification}
  \begin{tabular}{|r|ll|rr|}
    \hline
    \textbf{Index} & \textbf{Aug./Norm.\ Method} & {(Modification)} & \textbf{Acc.} & (std) \\ % & \textbf{AUC} (std) \\
\hline
\multicolumn{5}{|c|}{\textbf{Baseline and Prior Arts}} \\
\hline
    1 & (Baseline) \cite{eman22}  &   & 0.561 & {(0.020)} \\ % & 0.628 \\
% \multicolumn{4}{c}{\textbf{Prior Arts in Augmentation}} \\
    2 & Augmentation w/ Style Transfer \cite{eman22} &   & 0.761 & {(0.018)} \\ % & 0.719 \\
\hline
\multicolumn{5}{|c|}{\textbf{Proposed Augmentation and Normalization Methods}} \\
\hline
% Replacing all entries
    3 & Proposed Augmentation (\cref{equ:aug}) &   & 0.772 & (0.014) \\ % & 0.717 (0.019) \\
    4 & Proposed Normalization (\cref{equ:norm}) &   & 0.764 & (0.036) \\
\hline
\multicolumn{5}{|c|}{\textbf{Ablation Studies}} \\
\hline
% Replacing some entries
% Ours & Direct Sampler & 0.759 (0.037) & 0.710 (0.019) \\
5 & Proposed Augmentation (\cref{equ:aug}) & (w/ Flow Sampler) & 0.765 & (0.028) \\ % & 0.717 (0.019) \\
6 & Proposed Augmentation (\cref{equ:aug}) & (w/ Independent Sampler) & 0.759 & (0.014) \\ % & 0.717 (0.019) \\
% Ours & Flow Sampler & 0.765 (0.028) & 0.701 (0.019) \\
% Ours & Independent Sampler & 0.759 (0.014) & 0.707 (0.017) \\
    7 & Proposed Normalization (\cref{equ:norm}) & (w/o post-processing) & 0.727 & (0.009)  \\
    8 & Proposed Normalization (\cref{equ:norm}) & (w/ naive greyscale
    conversion) & 0.722 & (0.015) \\
    \hline
  \end{tabular}
\end{table}
\begin{figure}[t]
  \centering
  % \begin{subfigure}{0.48\textwidth}
  %   \includegraphics[width=\textwidth]{imgs/demos/quadratic.png}
  %   \caption{With Post-Processing}
  %   %\label{fig:first}
  % \end{subfigure}
  % \hfill
  \begin{subfigure}{0.48\textwidth}
    \includegraphics[width=\textwidth]{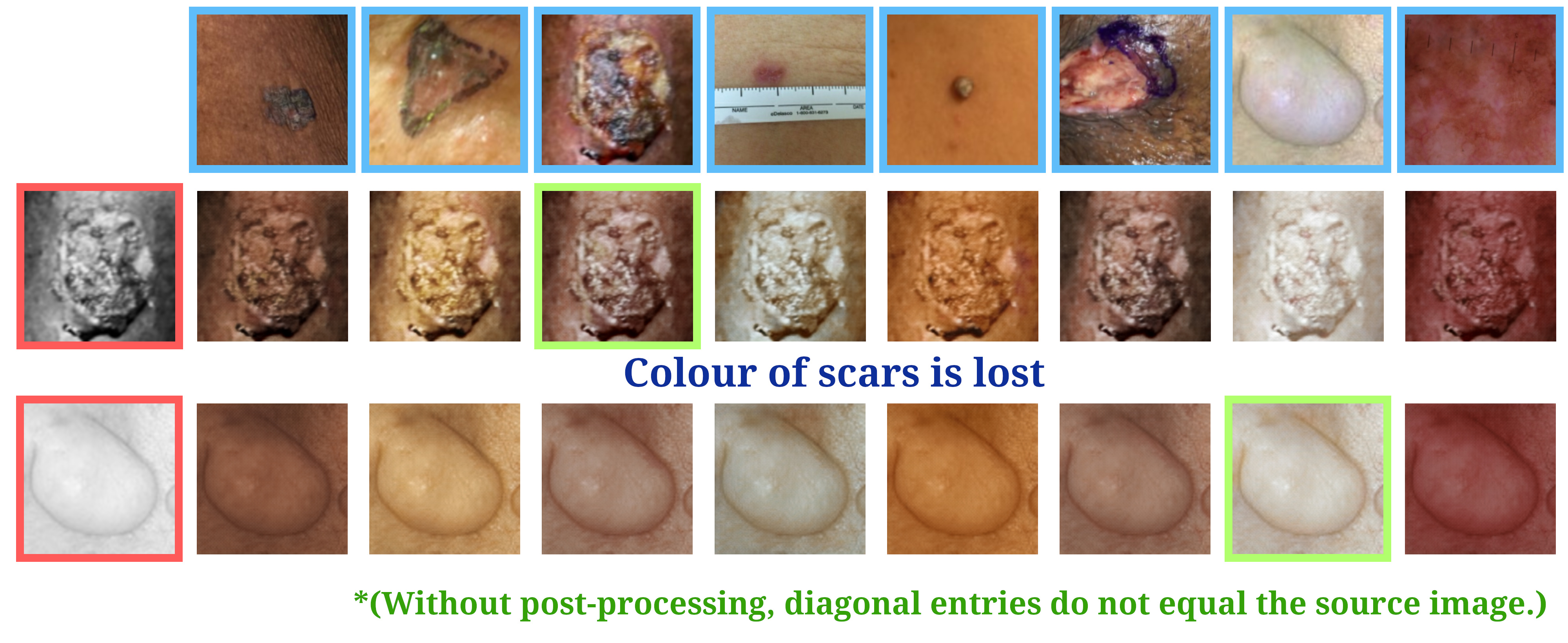}
    \caption{Without Post-Processing}
    \label{fig:no-pp}
  \end{subfigure}
  % \hfill
  \begin{subfigure}{0.48\textwidth}
    \includegraphics[width=\textwidth]{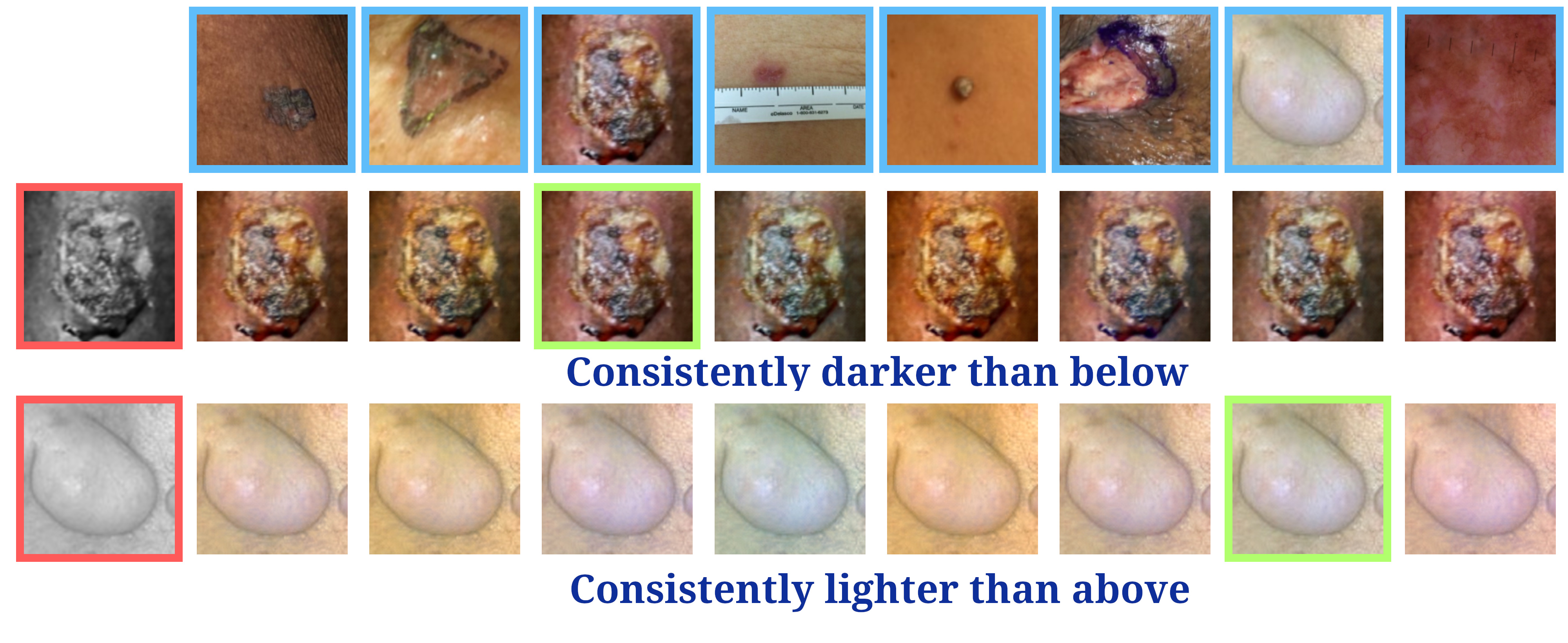}
    \caption{With naive decolourization method}
    %\label{fig:second}
  \end{subfigure}
  \caption{Qualitative ablation study results of 
    transferring perceived skin colours.  
    Rows and columns are defined in the same way as in
    \cref{fig:great}(a).
    % Each row indicates the source image for skin and lesion structure,
    % and each column indicates the source image for skin colour.  
    % The diagonal entries on the right are the reconstructed images and
    % those on the left equals the input images.
  }
  \label{fig:qualitative-transfer}
\end{figure}

Our method is a general framework applicable to various
downstream tasks.  Due to space constraints, we present only
the results of training a lesion malignancy classification
model.
Note that the goal of these experiments is not to propose a
state-of-the-art classifier, but to explore the potential of
our augmentation and normalization methods. % and to compare
%their effectiveness with existing approaches.
%
For fair comparison, we followed the same evaluation protocol as
in~\cite{eman22} and obtained the training set (parts of DermNet
NZ~\cite{dermnetnz} and ISIC-2018 JID editorial
images~\cite{isic2018}) and test set (parts of Dermatology
Atlas~\cite{dermatlas}) directly from the
authors. %,  are obtained  of~\cite{eman22}.
For each augmented or normalized training set, we train a
ResNet-50~\cite{resnet} model for lesion malignancy classification
using Adam with a learning
rate of \(10^{-3}\), repeating each experiment ten times and
reporting the mean and standard deviation of accuracy.
% and the area under the curve (AUC) on the test set.  
%
%
The colour model for data augmentation is trained on unlabelled training
and test sets for classification, while the colour normalization model is trained
on the training set only (to improve representation
quality, unlabelled
ISIC-2020~\cite{isic2020} is added to the set).
%Rows~\tobemodified{3} and~\tobemodified{4} in
%\cref{tbl:classification} show the performance of models
%trained on the augmented and normalized training sets,
%respectively.
The models trained on the augmented (row 3 in \cref{tbl:classification}) and normalized (row 4) sets
outperform the baseline
(row 1) by a large margin. % , with $p < 0.001$ (Welch's $t$-test).
% Augmentation outperforms ($p < 0.05$) the prior method ~\cite{eman22}
% (row 2), while the normalization method performs favorably comparable. % ($p < 0.25$).
They also perform favorably compared to the prior augmentation method ~\cite{eman22},
while providing significantly more interpretability in the representation space. 
% that our model is able to
% achieve improvements in accuracy. 
%  while providing significantly more interpretability in the augmentation process and representation space. 
% \todo{it might be necessary to add more 
%   ablation studies on the sampling methods, as the current methods'
%   performance are really close. 
% }

\subsection{Ablation Study}

\notebox{
Show the following qualitative results:

\begin{itemize}
  \item Effect of Decolourization
  \item Effect of Post-Processing
  \item Effect of bpp loss (only if we have space to discuss latent
    analysis)
  \item Latent sampling methods (to justify the simpler sampling
    method)
\end{itemize}
}

% \subsubsection{Effects of Post-Processing}
\textbf{Effects of Post-Processing}
% The network output without the post-processing procedure is shown in
% \cref{fig:qualitative-transfer}(a).  
% the results with post-processed (\cref{fig:great}(a)) and
% the one without (\cref{fig:qualitative-transfer}(a)) that, 
% (b) 
Certain non-skin features (e.g., scars and ink marks) that
are faithfully preserved after post-processing
(\cref{fig:great}(a)) are absent in the direct network output
(\cref{fig:qualitative-transfer}(a)).
When post-processing is removed from the normalization
pipeline, the downstream classifier performs significantly
worse than the full method
(row 7 vs.\ row 4 in
\cref{tbl:classification}).
We also computed the average LPIPS
score~\cite{lpips} between the normalized and original images
(after masking out non-lesion regions
using masks provided by~\cite{rezk2022leveraging}), and found that removing
post-processing worsens the score from 0.011 to 0.045.

% Normalization Results:
  % lpips_after: 0.011175
  % lpips_before: 0.044962
  % ssim_after: 0.989971
  % ssim_before: 0.955914

% This indicates the post-processing procedure's effectiveness in 
% effectively correct the colour shifts of such features, while still
% maintaining the transferred skin colour.

% \subsubsection{Effects of Decolourization Methods}
\textbf{Effects of Decolourization Methods}
The naive alternative to the proposed decolourization method is the 
default greyscale conversion method in PyTorch \cite{pytorch}.  
In contrast to \cref{fig:great}(a), 
the model trained with the naive method's outputs
produces results (\cref{fig:qualitative-transfer}(b)) whose skin darkness
resembles that of the target image (which should have provided lesion
structure) rather than the source image (which should have provided SCCI).  
This indicates the naive conversion is not capable of enabling faithful
skin colour transfer and can 
create significant
obstacle to provide diverse augmentation results for education.

% In addition to the proposed decolourization method, the default
% greyscale conversion method provided by PyTorch \cite{pytorch} is also
% tested (shown in \cref{fig:qualitative-transfer}(a)).  In contrast to 
% \cref{fig:great}(a), the darkness of skin is not properly
% transferred when using the default method, indicating the leakage of
% information.

% Normalization Results w/ default greyscale conversion:
  % lpips_after: 0.005891
  % lpips_before: 0.044139
  % ssim_after: 0.997902
  % ssim_before: 0.950193

% \subsubsection{Effects of Latent Sampling Methods}
\textbf{Effects of Latent Sampling Methods}
Under ideal conditions (a very large training set and a
highly flexible model), the entries of \(\bm e\) may be
assumed independent; in practice, however, weak correlations
between entries are observed.
We therefore compare several latent sampling strategies for
the augmentation method; results are shown at the bottom of
\cref{tbl:classification}.
% \todo{add Naive Gaussian and uniform}
We tested 3 methods for sampling the latent
vector \(\bm e\) from the distribution of the test set:
(row 3) directly reusing latents from the test set (by random
selection),
(row 5) fitting a normalizing flow model~\cite{realnvp}, and
(row 6) sampling each entry independently from the empirical
marginal distribution.
All methods with a relatively accurate modelling of
the marginal distribution of the latent vector can achieve comparable
performance, while direct random sampling performs the best.

% refer to \cref{fig:qualitative-transfer}(b) here

\section{Conclusion}

\notebox{
Also mention this tool can be used as a training tool for new medical 
professionals
}
In this work, we presented a disentanglement-by-compression framework that learns a
structured, manipulable latent representation of skin colour captured
in images (SCCI) from unlabelled dermatology data.
A randomized decolourization mapping and a geometry-aligned
post-processing step were proposed to improve faithfulness of the
colour manipulation; ablation studies confirmed that both contribute
measurably to downstream performance and perceptual fidelity.
The learned latent space supports colour transfer, independent entry
manipulation, and traversal along physically meaningful
trajectories (e.g., blood perfusion, camera white balance),
enabling training set augmentation, colour normalization, and
educational visualization of skin conditions under varying SCCI.
On a downstream benchmark, both augmentation
and normalization strategies perform favourably against
% the unaugmented baseline and a prior method.
prior methods. 
This work is a step towards achieving equitable diagnosis, where our models can be reliably adapted across different clinical settings and populations. 
\bibliographystyle{splncs04}
\bibliography{main.bib}
\end{document}